\documentstyle[prl,aps,epsf,multicol]{revtex}
\begin{document}
\title{\bf Exact Solution of a Drop-push Model for Percolation}

\author{Satya N. Majumdar and David S. Dean}

\address{Laboratoire de Physique Quantique (UMR C5626 du CNRS),
Universit\'e Paul Sabatier, 31062 Toulouse Cedex, France. \\}

\date{\today}

\maketitle

\begin{abstract}
Motivated by a computer science algorithm known as `linear probing
with hashing' we study a new type of percolation model whose basic
features include a sequential `dropping' of particles on a substrate
followed by their transport via a `pushing' mechanism. Our exact
solution in one dimension shows that, unlike the ordinary random
percolation model, the drop-push model has nontrivial spatial
correlations generated by the dynamics itself.  The critical exponents
in the drop-push model are also different from that of the ordinary
percolation.  The relevance of our results to computer science is
pointed out.

\noindent

\end{abstract}

\begin{multicols}{2}

The ordinary site or bond percolation and its various generalizations are
amongst the most well studied problems in statistical
physics\cite{SA,physica}.  Motivated by a well known computer science
algorithm known as `linear probing with hashing' (LPH)\cite{Knuth}, we
introduce and study in this Letter a new type of percolation
model. Borrowing a name from the models of activated flow through
traps\cite{BR} we call this a `drop-push' model since it has two basic
features: a sequential `dropping' of particles on a substrate followed
by the transport of the dropped particles via a `pushing' mechanism
caused by the local hardcore repulsion between particles on the
substrate.  Unlike in the ordinary percolation, we show that the 
dynamics of the drop-push model generates
nontrivial spatial correlations between sites.  Our exact solution in one dimension
shows that the critical exponents associated with the percolation
transition in the drop-push model are different from those of 
the ordinary percolation.  As an additional bonus, our approach also
rederives, in a straightforward way, a key result on the cost function
in the LPH algorithm obtained recently by computer scientists
using more involved combinatorial techniques\cite{FPV,Janson}.  Our
model is also easily generalizable to higher dimensions.

The LPH algorithm was originally introduced by Knuth\cite{Knuth} and
has remained popular in computer science due to its simplicity,
efficiency and general applicability\cite{FPV}.  It can be described
as follows: Consider $M$ items $x_1$, $x_2$, $\ldots$, $x_M$ to be
placed sequentially into a table with $L$ cells labelled $1$, $2$,
$\ldots$, $L$.  For each item $x_i$, a cell with label $h_i\in
\{1,2,\ldots ,L\}$ is selected. The label $h_i$ is called the hash
address and is usually chosen randomly from the set $\{1,2,\ldots,
L\}$. The item $x_i$ is inserted at the $h_i$-th cell provided it is
empty. Otherwise one tries cells $h_i+1$, $h_i+2$, etc.  until an
empty cell is found (the locations of the cells are interpreted modulo
$L$) where the item $x_i$ is finally inserted. From the computer
science point of view, the object of interest is the cost function
$S(M,L)$ defined as the total number of unsuccessful probes
encountered in inserting the $M$ items into a table of size
$L$\cite{FPV,Janson}. This cost function measures the construction
cost of the table as well as the time to search for an item
later\cite{FPV,Janson}.  The statistics of this cost function were
found to be very different in the sparse table (when the density $M/L
<<1$) compared to the full table (when $M/L\to 1$)\cite{FPV}.  We show
below that, when interpreted as an interacting particle system, this
crossover from the sparse to full table corresponds precisely to a
percolation transition which belongs to a different universality class
than that of the ordinary site or bond percolation.
\begin{figure}
  \narrowtext\centerline{\epsfxsize\columnwidth \epsfbox{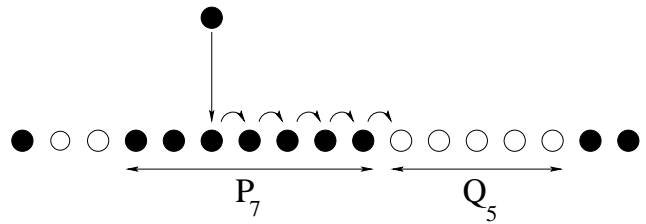}}
\caption{The dropping and the subsequent hopping moves in the
drop-push model. Also shown a particle cluster of size $7$ and a hole
cluster of size $5$.}
\end{figure}

In our equivalent drop-push model (Fig. 1), we interpret the table as
a lattice of size $L$ with periodic boundary conditions. The cells are
the lattice sites and each site can contain at most one particle. One
starts with an empty lattice. At each step a particle is dropped into
a randomly selected site.  If the target site is empty, the incoming
particle occupies it. If however the site was already occupied, the
particle keeps hopping to the right until it finds an empty site 
where it then stays (see Fig. 1).  One then repeats the same procedure 
with the
next particle and so on. The dropping process in this model is similar
to that of random sequential adsorption (RSA) and the car parking
processes\cite{RSA}, however the adsorption mechanism in the drop-push
model is quite different from the usual RSA models.  In  the car
parking language, in this drop-push model, a new car arrives at a random
spot on a one way lane and moves forward till it finds an empty
parking spot.  Unlike usual car parking models, here a car always
manages to find a spot and thus the system never gets stuck in a
jammed state.  Note that although we have defined the hopping of the
particle to be unidirectional, one can also consider an unbiased
version where a particle, if dropped onto an occupied cluster,
performs an unbiased random walk on the occupied cluster till it finds
an empty site where it is then attached. We show below that in one
dimension, the results are independent of this bias.

Each addition of a new particle corresponds to incrementing the
density $t=M/L$ by the amount $\Delta t =1/L$.  Thus in the
thermodynamic limit $L\to \infty$, the density $t$ becomes a
continuous `time' like variable that increases monotonically from
$t=0$ (empty lattice) to $t=1$ (full lattice). For convenience, we
will henceforth refer to the density $t$ as `time' with $0\le t\le 1$.
The central objects of our analysis are $P_n(t)$ and $Q_n(t)$ denoting
respectively the number (per unit length) of particle and hole clusters
of size $n$ at time $t$ (see Fig. 1).  The total number of particle
(hole) clusters is denoted by $N(t)=\sum_n P_n(t)=\sum_n Q_n(t)$.
Note that as $t$ increases continuously from $0$ to $1$, one expects
that $N(t)$ starting from $N(0)=0$ should increase initially but will
eventually decrease to $0$ again as $t\to 1$ when the lattice is
nearly full. Thus $N(t)$ has an interesting nonmonotonic behavior in
$0\le t\le 1$ with a maximum (when the system has the largest number
of clusters) at an intermediate time $t^{*}$ (see the inset of Fig. 2).  We
also note the sum rules corresponding to the particle and the hole
densities: $\sum_n nP_n(t) = t$ and $\sum_n nQ_n(t)=(1-t)$.
\begin{figure}
  \narrowtext\centerline{\epsfxsize\columnwidth \epsfbox{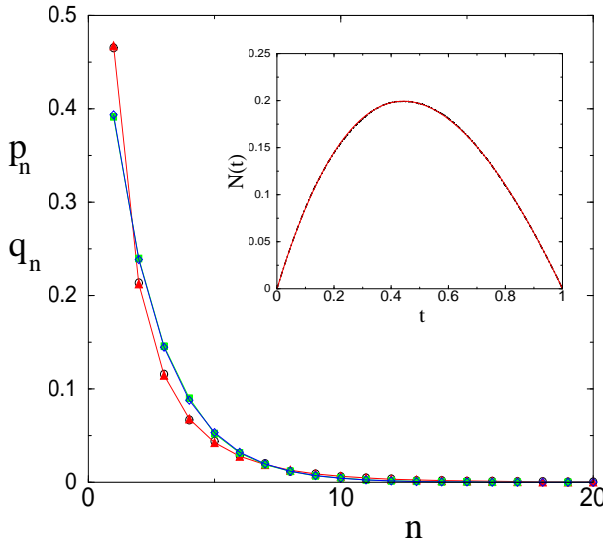}}
\caption{The exact solutions for the normalized cluster densities
$p_n=P_n/N$ (filled triangles) and $q_n=Q_n/N$ (filled squares) are
compared with numerical results (circles and diamonds
respectively) obtained via the Monte Carlo simulation of the drop-push
model on a lattice with $L=100000$ at the filling factor $t=0.5$.  The
inset shows a plot of the numerical domain density $N(t)$ and the
exact result $N(t)=(1-t)(1-e^{-t})$. The two curves are
indistinguishable.}
\end{figure}

The key observation that allows the exact solution in one dimension is
the fact that the `cluster' or `domain' densities $P_n$'s and $Q_n$'s
are statistically independent at all times $0\le t \le 1$.  This
follows from the fact that the adsorption of a new particle at the
edge of a particle (hole) cluster does not introduce correlations
between the adjacent particle and hole clusters. A rigorous
justification of this fact, details of which will be published
elsewhere\cite{MD}, follows from the observation that the domain walls
in the drop-push model can be viewed as the zero crossings of a Markov
process in space (at a fixed time $t$). Thus if one labels a
configuration $\cal C$ by the set $\{n_i\}$ where $n_i$'s denote the
lengths of the alternate particle and hole clusters, then the
probability of $\cal C$ is given by the product measure, ${\rm
Prob}[{\cal C},t]\propto
P_{n_1}(t)Q_{n_2}(t)P_{n_3}(t)Q_{n_4}(t)\ldots$. In other words, the
independent interval approximation (IIA) is exact in this model. The
next step is to write down the exact rate equation of evolution of
$P_n$'s and $Q_n$'s by accounting for all the gain and loss terms due to
the addition of a new particle and exploiting the factorization
property of the probabilities. Similar types of IIA equations have been 
used in  one dimensional coarsening problems\cite{KB}. The
rate equation for the $Q_n$'s turn out to be simple
\begin{equation}
{{dQ_n}\over {dt}}= -\left(n+ {t\over {N}}\right)Q_n +
2\sum_{m=n+1}^{\infty} Q_m +{t\over {N}}Q_{n+1},
\label{qe}
\end{equation}    
valid for all $n\ge 1$.  The first term denotes the loss of a hole
cluster of size $n$ due to an adsorption of a particle by direct hit
at any of the available $n$ sites and also due to a hit at any
occupied site of the neighboring particle cluster to the left which
then transports the particle to the leftmost site of the hole
cluster. The latter happens with rate $\sum_k kP_k/N=t/N$.  The second
term denotes a gain in $Q_n$ due to a direct hit inside a hole cluster
of size bigger than $n$.  The factor $2$ denotes that there are only
two places available for the incoming particle in order to generate a
hole cluster of size $n$ from a bigger hole cluster. The third term
indicates a gain of a hole cluster of size $n$ from that of size $n+1$
due to a particle adsorption at its left edge.  One can similarly
write down the equations for the $P_n$'s though they turn out to be
trickier. Omitting the details\cite{MD}, we present only the final results
\begin{eqnarray}
{{dP_n}\over {dt}}&=& -\left(n+2+{{tQ_1}\over {N^2}}\right)P_n
+\left(1-{{Q_1}\over {N}}\right)(n+1)P_{n-1}\nonumber \\
&+&{{Q_1}\over {N^2}}\sum_{j=1}^{n-2}(j+1)P_j P_{n-1-j}, \,\,\, n\ge 2
\nonumber \\ {{dP_1}\over {dt}}&=& -\left(3 +{{tQ_1}\over
{N^2}}\right)P_1 + \sum_{m=2}^{\infty} (m-2)Q_m.
\label{pe}
\end{eqnarray}
Although the above equations are written down for the unidirectional
version of the model, a careful analysis shows that they remain
unchanged for the general case in which the dropped particle moves
to the right with probability $p$ or to the left with probability
$(1-p)$\cite{MD}.

As an important consistency check, one can verify that both the
Eqs. (\ref{qe}) and (\ref{pe}) satisfy the respective sum rules
$\sum_n nQ_n =(1-t)$ and $\sum_n nP_n =t$. One can also write down the
evolution equation for the total domain density via direct inspection
of the process,
\begin{equation}
{{dN}\over {dt}}=-2 N - {t\over {N}}Q_1 + 1-t.
\label{Ne}
\end{equation}
It is easy to check that both  Eqs. (\ref{qe}) and (\ref{pe}) when
summed over $n$ satisfy  Eq. (\ref{Ne}) thus providing yet another
useful consistency check.

We note that Eq. (\ref{qe}) for the $Q_n$'s does not involve the
$P_n$'s, but is however implicitly nonlinear due to the occurrence of
$N=\sum_n Q_n$ on the right hand side. However Eq. (\ref{qe})
admits a very simple pure exponential solution as found in many  RSA
models\cite{RSA}. The ansatz $Q_n(t)=A(t)\exp[-n B(t)]$ satisfies
Eq. (\ref{qe}) for all $n\ge 1$ with the choice $A(t)=
2(1-t)[\cosh(t)-1]$ and $B(t)=t$. One then gets the exact solution for
the domain density $N(t)=\sum_n Q_n(t)= (1-t)(1-e^{-t})$ which is
nonmonotonic in $0\le t \le 1$ and is asymmetric about its unique
maximum at $t^{*}=0.4428\ldots$ (see Fig.  2). Note that in the ordinary
site percolation in $1$-d with occupation probability $t$, the domain
density is simply given by $N(t)=t(1-t)$ which is symmetric about the
maximum at $t^*=1/2$.

We next substitute the exact solution for the $Q_n$'s into
Eq. (\ref{pe}) and first solve for $n=1$ and $n=2$. The exact
solutions $P_1(t)= t(1-t)e^{-2t}$ and $P_2(t)=3t^2(1-t)e^{-3t}/2$
suggest the ansatz: $P_n(t)=a_n(1-t)t^n e^{-(n+1)t}$. Indeed this
ansatz solves Eq. (\ref{pe}) provided the $a_n$'s satisfy the
nonlinear recursion relation,
\begin{equation}
na_n = (n+1)a_{n-1} +\sum_{i=1}^{n-2} (i+1)a_ia_{n-1-i},
\label{ae}
\end{equation}
starting with $a_1=1$. The first few values are $a_2=3/2$, $a_3=8/3$,
$a_4=125/24$ etc.  From Eq. (\ref{ae}) it follows that the generating
function ${\tilde a}(z)=\sum_{n=1}^{\infty} a_n z^n$ is given by
${\tilde a}(z)= -1+ T(z)/z$ where $T(z)$ is the well known tree
function given by the solution of the equation $T(z)\exp[-T(z)]=z$. It
is easy to see that the function $T(z)$ has a singularity at $z=1/e$
where $T(z)\approx 1- \sqrt{2(1-ez)} +\ldots$. This tree function
appears in problems related to the counting of rooted labelled
trees\cite{GJ} with various applications in computer
science\cite{Knuth,FPV} as well as many physics problems such as
aggregation models\cite{DE} and the classical hard sphere fluid in large
dimensions\cite{FRW}.  Using the known properties of $T(z)$, we then
get $a_n= (n+1)^{n-1}/n!$ for all $n\ge 1$.  We have also checked that
the exact solutions
\begin{eqnarray}
Q_n(t)&=& (1-t){\left(e^t-1\right)}^2 \, e^{-(n+1)t} \\ P_n(t)&=&
(1-t)t^n e^{-(n+1)t} {{(n+1)^{n-1}}\over {n!}}
\label{pqs}
\end{eqnarray}
match perfectly with the numerical results obtained via the Monte
Carlo simulation of the drop-push model (see Fig. 2).

Clearly at $t=1$, there is only one infinite particle cluster and the
system percolates.  We now analyze the scaling behavior of the cluster
distributions near the critical point $t=1$. From Eq. (\ref{pqs}), we
find that for large $n$, $P_n(t)\approx {{(1-t)}\over
{\sqrt{2\pi}}}n^{-\theta}\exp[-n/n^*(t)]$ where $n^*(t)=1/(t-1-\log
t)$ and the Parisi-Sourlas exponent\cite{PS} is given exactly by
$\theta=3/2$. Note that for the ordinary percolation in one dimension,
$P_n(t)=(1-t)^2 t^n $ indicating $\theta=0$. In the limit $t\to 1$,
the typical cluster size diverges as $n^*(t)\approx 2(1-t)^{-2}$ and
one obtains the Stauffer scaling form\cite{SA} for the cluster size,
$P_n(t)\sim n^{-\tau}f[n(1-t)^{\sigma}]$ with the Fisher exponent
$\tau=2$ (as in the ordinary percolation) and $\sigma=2$ (in contrast to
the ordinary percolation where $\sigma=1$). The exact scaling function
here $f(z)=\sqrt{{z}\over {2\pi}}e^{-z/2}$ also differs from that of
the ordinary percolation where $f(z)=z^2e^{-z}$\cite{SA}.  Consequently the
susceptibility exponent\cite{SA} given by the scaling relation
$\gamma= (3-\tau)/\sigma$ also differs in the two models. For the
drop-push model $\gamma=1/2$, where as   $\gamma=1$ for 
the ordinary percolation.

We now turn to the correlation function $G_n(t)$ denoting the
probability that two occupied sites separated by a distance $n$ belong
to the same cluster at time $t$. For convenience we introduce the
binary variable $\sigma_i$ such that $\sigma_i=1$ if the site $i$ is
occupied and $\sigma_i=0$ otherwise. Then by definition $G_n=\langle
\sigma_i \sigma_{i+1} \ldots \sigma_{i+n}\rangle $.  We also note that
by definition the particle cluster density $P_n =\langle {\bar
\sigma_i} \sigma_{i+1}\ldots \sigma_{i+n} {\bar
\sigma_{i+n+1}}\rangle$ where ${\bar \sigma_i}=1-\sigma_i$. As a
consequence one obtains the exact relation,
$G_{n+1}+G_{n-1}-2G_n=P_n$. Using the exact scaling form of $P_n$ and
integrating twice with respect to $n$ we find that in the scaling
limit $t\to 1$, $n\to \infty$, $G_n(t)\approx g[n(1-t)^\nu]$ where the
correlation length exponent $\nu=2$ and the exact scaling function
$g(z)= (1+z){\rm erfc} (\sqrt{{z\over 2}})-\sqrt{ {{2z}\over
{\pi}}}e^{-z/2}$.  These results should be compared to those for the
ordinary percolation with occupation density $t$ where $G_n(t)=t^n$
indicating that $\nu=1$ and $g(z)=e^{-z}$ trivially.

To elucidate the nontrivial spatial correlations in the drop-push
model we have also computed the conventional two point correlation function,
$C_n(t)=\langle \sigma_i \sigma_{i+n}\rangle -\langle \sigma_i \rangle
\langle \sigma_{i+n}\rangle$, the connected part of the
joint probability that two sites at distance  $n$  are both occupied.
Note that for the ordinary percolation, $C_n(t)=0$ for all $n\ge 1$
and $0\le t\le 1$. In contrast, we show that $C_n(t)$ is
nontrivial in the drop-push model. In order to compute it, we 
add up the possibilities that there may be no holes between the two
sites, or maybe only one hole cluster, or two hole clusters etc. This
method of expressing the correlation function in terms of the interval
size distributions was used before in other contexts\cite{AB,diffusion}. 
Omitting details we present only
the final expression for the generating function ${\tilde C}(z)=\sum_n
C_n(t) z^n$
\begin{equation}
{\tilde C}(z)={{1-t}\over {1-z}}\left[zt-1+{{t(1-z)}\over
{t-T(zte^{-t})}}\right],
\label{Cz}
\end{equation}
where $T(z)$ is the tree function defined earlier. Using the
properties of the tree function it is straightforward, though
somewhat tedious\cite{MD}, to derive the asymptotic scaling properties of
$C_n(t)$. We find after some algebra that in the scaling limit $t\to
1$, $n\to \infty$ but keeping $z=n(1-t)^2$ fixed, $C_n(t)\approx
(1-t)^2 F[n(1-t)^2]$ where the exact scaling function $F(z)={1\over
{\sqrt{ 2\pi z}}}e^{-z/2} -{1\over {2}}{\rm erfc}(\sqrt{{z\over
2}})$. The function $F(z)\approx 1/{\sqrt {2\pi z}}$ as $z\to 0$ and
$F(z)\approx {1\over {\sqrt{2\pi}z^{3/2}}}e^{-z/2}$ as $z\to \infty$.

We now turn to the total cost function $S(t)$ in the LPH algorithm,
i.e.  in the drop-push model with unidirectional transport. The cost
$\Delta S(t)$ to insert a new particle at time $t$ is precisely the
expected number of unsuccessful probes, i.e., the expected number of
hops that the new particle undergoes before getting adsorbed into the
right edge of a particle cluster.  Consider a particle cluster of size
$k$ at time $t$. The incoming particle can drop anywhere on this
cluster and the number of hops is simply the distance of the target
site from the right edge of the cluster. Noting that the cluster size
can vary from $1$ to infinity, one then gets $\Delta S(t)
=\sum_{k=1}^{\infty}\left(\sum_{j=1}^k j\right)P_k(t)={1\over
2}\sum_{k=1}^{\infty}k(k+1)P_k(t)$.  Using the exact result for
$P_k(t)$ from Eq. (\ref{pqs}) we get $\Delta S(t)=
t(2-t)/[2(1-t)^2]$. The total cost function is then given by
$S(t)=\int_0^{t}\Delta S(t')dt'=t^2/[2(1-t)]$, in agreement with the
result derived by the computer scientists using rather involved combinatorial
techniques\cite{FPV,Janson}.  More details on the statistics of this
cost function for finite size tables have also been derived
recently\cite{FPV,Janson}.

Finally this drop-push model can be easily generalized to
higher dimensions. Let us consider, for simplicity, the 
the unbiased version.  One starts with an empty lattice
of linear size $L$ in $d$-dimensions with periodic boundary
conditions. At each step one drops a particle at a randomly chosen
site. The incoming particle occupies the target site provided it was
empty. Otherwise the dropped particle performs a random walk starting
at the target site until it finds an empty site and sticks there.
One then adds another particle and the process continues. It follows
from the simple electrostatic analogy that when a particle drops onto
an occupied cluster, it has equal probability to subsequently stick to
any of the surface sites of this cluster. In this sense this model to
similar to the celebrated Eden model\cite{Sander}.  However, unlike the
Eden model, here one can have many different seed sites
from which a new cluster can grow. Also the
probability that a given cluster will grow by absorbing a new particle
is proportional to the volume of the cluster in the drop-push model.
Note that this model is also different from the previously studied
cooperative adsorption models\cite{SE}, the cluster-cluster
aggregation models\cite{Sander} and the random dynamical percolation
model\cite{FLR}. It is clear that there will be a critical density
$t=t_c<1$ at which the particle clusters start to percolate.  An
outstanding question is whether this percolation transition in higher
dimensions, as in the $1$-d case, belongs to a different universality
class than that of the ordinary site percolation. Our preliminary numerical
simulations in $2$-d indicate that indeed this may be the case\cite{MD}. We
defer these results and other details for a future communication.

We thank M. Barma and D. Dhar for useful discussions.

\end{multicols}


\begin{references}

\bibitem{SA} D. Stauffer and A. Aharony, {\em Introduction to
Percolation Theory}, 2nd ed. (Taylor and Francis, London, 1992).

\bibitem{physica} For various generalizations of the ordinary
percolation see the articles in the special issue, Physica A {\bf 266}
(1999).


\bibitem{Knuth} D.E. Knuth, {\em The Art of Computer Programming}, 2nd
ed. Vol-3 (Addison-Wesley, Reading, 1998).

\bibitem{BR} M. Barma and R. Ramaswamy in {\em Nonlinearity and
Breakdown in Soft Condensed Matter}, edited by K.K. Bardhan,
B.K. Chakrabarti, and A. Hansen (Springer, Berlin, 1993).

\bibitem{FPV} P. Flajolet, P. V. Poblete, and A. Viola, Algorithmica,
{\bf 22}, 490 (1998).

\bibitem{Janson} S. Janson, Random Struct. Alg. {\bf 19}, 438 (2001).
                                                                     
\bibitem{RSA} M.C. Bartelt and V. Privman, Int. J. Mod. Phys. {\bf
B5}, 2883 (1991); J.W. Evans, Rev. Mod. Phys. {\bf 65}, 1281 (1993);
J. Talbot, G. Tarjus, P.R. Van Tassel, and P.  Viot, Colloid Surface A
{\bf 165}, 287 (2000).

\bibitem{MD} S.N. Majumdar and D.S. Dean, unpublished.

\bibitem{KB} P.L. Krapivsky and E. Ben-Naim, Phys. Rev. E {\bf 56},
3788 (1997); S.N. Majumdar, D.S. Dean, and P. Grassberger,
Phys. Rev. Lett. {\bf 86}, 2301 (2001).

\bibitem{GJ} I. P. Goulden and D.M. Jackson, {\em Combinatorial
Enumeration} (John Wiley, New York, 1983).

\bibitem{DE} P.G.J. van Dongen and M.H. Ernst, J. Phys. A:
Math. Gen. {\bf 16}, L327 (1983); J.L. Spouge, J. Phys. A:
Math. Gen. {\bf 16}, 767 (1983).

\bibitem{FRW} H.L. Frisch, N. Rivier, and D. Wyler,
Phys. Rev. Lett. {\bf 54}, 2061 (1985).

\bibitem{PS} G. Parisi and N. Sourlas, Phys. Rev. Lett. {\bf 46}, 871
(1981).

\bibitem{AB} P.A. Alemany and D. ben-Avraham, Phys. Lett. A {\bf 206}, 18 (1995).

\bibitem{diffusion} S.N. Majumdar, C. Sire, A.J. Bray, and
S.J. Cornell, Phys. Rev. Lett. {\bf 77}, 2867 (1996); B. Derrida,
V. Hakim, and R. Zeitak, {\em ibid.} {\bf 77}, 2871 (1996).

\bibitem{Sander} For a review of cluster growth models see L.M. Sander
in {\em Solids Far from Equilibrium}, edited by C. Godr\`eche
(Cambridge University Press, Cambridge 1992).

\bibitem{SE} D.E. Sanders and J.W. Evans, Phys. Rev. A {\bf 38}, 4186
(1988).

\bibitem{FLR} J.E. de Freitas, L. dos Santos Lucena, and S. Roux,
Physica A, {\bf 266}, 81 (1999).


\end{references}
\end{document}